\documentclass[preprint,3p,sort&compress]{elsarticle}

\usepackage{amsmath,amssymb}
\usepackage{epsfig}
\usepackage{bbm}
\usepackage[pdftex]{hyperref}

\begin{document}
\vspace*{-0.3cm}
\hspace*{13cm} \hbox{\rm YITP-SB-11-08}
\vskip 1cm
\title{GRBs on probation: testing the UHE CR paradigm with IceCube}
\author[sb]{M.~Ahlers} 
\ead{ahlers@insti.physics.sunysb.edu}
\author[sb,ub]{M.~C.~Gonzalez--Garcia}
\ead{concha@insti.physics.sunysb.edu}
\author[uw]{F.~Halzen}
\ead{halzen@icecube.wisc.edu}
\address[sb]{%
  C.N.~Yang Institute for Theoretical Physics,
  SUNY at Stony Brook, Stony Brook, NY 11794-3840, USA}
\address[ub]{
  Instituci\'o Catalana de Recerca i Estudis Avan\c{c}ats (ICREA),
  Departament d'Estructura i Constituents de la Mat\`eria and ICC-UB, 
  Universitat
  de Barcelona, 647 Diagonal, E-08028 Barcelona, Spain}
\address[uw]{%
Department of Physics, University of Wisconsin, Madison, WI 53706,USA}

\begin{abstract}
Gamma ray burst (GRB) fireballs provide one of very few astrophysical
environments where one can contemplate the acceleration of cosmic rays
to energies that exceed $10^{20}$\,eV. The assumption that GRBs are
the sources of the observed cosmic rays generates a calculable flux of
neutrinos produced when the protons interact with fireball
photons. With data taken during construction IceCube has already
reached a sensitivity to observe neutrinos produced in temporal
coincidence with individual GRBs provided that they are the sources of
the observed extra-galactic cosmic rays. We here point out that the GRB
origin of cosmic rays is also challenged by the IceCube upper limit on
a possible diffuse flux of cosmic neutrinos which should not be
exceeded by the flux produced by all GRB over Hubble time. Our
alternative approach has the advantage of directly relating the
diffuse flux produced by all GRBs to measurements of the cosmic ray
flux. It also generates both the neutrino flux produced by the sources
and the associated cosmogenic neutrino flux in a synergetic way.
\end{abstract} 


\maketitle

\section{Motivation}
  
There is compelling evidence that the collapse of a massive star to a
black hole is the primary engine of long-duration Gamma Ray Bursts
(GRBs). The phenomenology that successfully accommodates the
astronomical observations is that of the creation of a hot fireball of
electrons, photons and protons that is initially opaque to
radiation. The hot plasma therefore expands by radiation pressure and
particles are accelerated to a Lorentz factor $\Gamma$ that grows
until the plasma becomes optically thin and produces the GRB
display. From this point the fireball is coasting with a Lorentz
factor that is constant and depends on its baryonic load. The baryonic
component carries the bulk of the fireball's kinetic energy. The
energetics and rapid time structure of the burst can be successfully
explained by shocks generated in the expanding
fireball~\cite{Shemi:1990rv,Rees:1992ek,Meszaros:1993tv}. Here, the
temporal variation of the $\gamma$-ray burst of the order of
milliseconds can be interpreted as the collision of internal shocks
with a varying baryonic load leading to differences in the bulk
Lorentz factor. Electrons accelerated by first order Fermi
acceleration radiate synchrotron $\gamma$-rays in the strong internal
magnetic field and thus produce spikes in the burst spectra of the
order of seconds~\cite{Rees:1994nw,Paczynski:1994uv}. The collision of
the fireball with interstellar gas forms external shocks that can
explain the GRB afterglow ranging from X-ray to the
optical~\cite{Meszaros:1993ft,Meszaros:1994sd}. (See
also~\cite{Meszaros:2006rc} for a recent review on theoretical
models.).

It has been pointed out that fireball baryons may be the source of
ultra-high energy (UHE) cosmic rays (CRs) with energies extending to
at least $3\times10^{20}$~eV. Baryons are inevitably accelerated along
with the electrons in the expanding fireball with their energies
boosted by the bulk Lorentz factor. It has been shown that a typical
GRB environment can naturally satisfy the requirements to produce UHE
CRs, most likely protons. It is not easy to conceive of a
mechanism where nuclei survive acceleration in a GRB fireball
environment. (Heavy nuclei could be synthesized in
magnetically-dominated jets as in the proto-magnetar model of GRBs,
see {\it e.g.}~\cite{Metzger:2011xs}.) The results of CR observatories disagree with
HiRes~\cite{Abbasi:2009nf} claiming protons and
Auger~\cite{Abraham:2010yv} heavier primaries. These conflicting
results may just illustrate that, given the poor knowledge of hadronic
interactions more than one order of magnitude above LHC energies, is
not sufficiently known to derive a definite
result~\cite{Ulrich:2009yq}.

Though GRBs satisfy the necessary conditions for accelerating
protons to UHE, it is problematic how these protons may eventually be
ejected as CRs: protons are magnetically confined to the expanding
fireball and its adiabatic cooling will reduce the maximum proton
energy significantly~\cite{Rachen:1998ir,Rachen:1998fd}. However, this
does not concern neutrons that are frequently produced in
$p\gamma$-interactions of accelerated protons with fireball photons by
 processes like $p\gamma \to \Delta^+ \to n\pi^+ $. Cosmic ray
protons could thus be identified as neutrons from
$p\gamma$-interactions that can escape from the magnetic environment
and $\beta$-decay back to protons at a safe distance. A smoking-gun
test of this scenario is the production of PeV neutrinos from the
decay of the charged pions inevitably produced along with the neutrons
~\cite{Waxman:1997ti,Bahcall:1999yr,Murase:2005hy,Anchordoqui:2007tn}. This neutrino flux
is in reach of large-scale neutrino telescopes like
IceCube~\cite{Ahrens:2003ix,Abbasi:2011qc}. (Possible emission of
neutrinos associated to the GRB progenitor and prior to the
$\gamma$-ray burst have been discussed
in~\cite{Razzaque:2002kb}. Neutrinos from proton interactions in
late internal or external shocks during the afterglow phase have been considered
in~\cite{Murase:2006dr,Murase:2007yt,Razzaque:2009zz}.)

In this paper we calculate the diffuse flux of neutrinos produced in
association with GRB cosmic rays by directly fitting the proton
spectra from the decay of fireball neutrons to CR data from HiRes. As
a byproduct we also obtain the flux of so-called GZK neutrinos
produced when the cosmic rays interact with microwave and
infro-red/optical background
photons~\cite{Greisen:1966jv,Zatsepin:1966jv}. The accompanying photon
flux resulting from the electromagnetic cascading of the neutral pions
peaks in the MeV-GeV energy region; we verify that it does not exceed
the extra-galactic diffuse $\gamma$-ray flux inferred by
Fermi-LAT~\cite{Abdo:2010nz}. Our main conclusion is that the
predicted flux exceeds the upper bound on a diffuse flux of cosmic
neutrinos obtained by IceCube from a year of data taken with half the
instrument during construction. Facing this negative conclusion, we
subsequently investigate the dependence of the predicted neutrino flux
on the cosmological evolution of the sources as well as on the
parameters describing the fireball. Although the latter are
constrained by the electromagnetic observation as well as by the the
requirement that the fireball must accommodate the observed cosmic ray
spectrum, the predictions can be stretched to the point that it will
take 3 years of data with the now completed instrument to conclusively
rule out the GRB origin of UHE CRs.

We work throughout in natural Heaviside-Lorentz units with
$\hbar=c=\epsilon_0=\mu_0=1$, $\alpha=e^2/(4\pi)\simeq1/137$ and
$1~{\rm G} \simeq 1.95\times10^{-2}{\rm eV}^2$.

\section{Fireball Model}

In the fireball model the GRB central engine produces a heated
optically thick plasma of leptons, photons and baryons which is
initially at rest. The fireball expands adiabatically by radiation
pressure until it becomes optically thin. From this point on the
fireball is coasting with a Lorentz factor $\Gamma$ which depends on
its baryonic load. The expanding plasma flow is shocked producing
shells with varying baryonic load; this results in internal shells
with different velocities that in their collision produce internal
shocks. Although in the baryon-rich fireballs that would produce EHE
cosmic rays most of the kinetic energy of the fireball is carried by
the baryons, the internal shocks will convert a fraction
$\varepsilon_e = U'_e/U'_{\rm kin}$ to leptons and $\varepsilon_B =
U'_B/U'_{\rm kin}$ to magnetic fields supported by the plasma. Here
and in the following, primed quantities refer to values in the
comoving plasma frame, whereas unprimed quantities are reserved for
the observer's frame.

Study of the emission of GRB has resulted in ``benchmark'' parameters
describing GRB fireball with the efficiency for lepton acceleration
from bulk energy taken to be 10\%, {\it i.e.}~$\varepsilon_e =
0.1$. The fraction of magnetic field energy and hence the value of the
turbulent magnetic field strength can be inferred from the observed
peak of the $\gamma$-ray synchrotron spectrum with $\varepsilon_B =
0.1$, corresponding to equipartition of energy in leptons and magnetic
field. The ``benchmark'' parameters are required to produce the
observed peak emission at MeV energies resulting from the synchrotron
radiation of GRB electrons and must be consistent with afterglow
observations. The typical radius of on internal shock $r_i$ is given
by the speed-of-light distance implied by the duration of the spikes
observed in the GRB display. Because the shells are boosted toward the
observer by a Lorentz factor $\Gamma_i$, $r_i \simeq
t_v/(1-\beta_i)\simeq 2\Gamma_i^2t_v$. We will take the variability
scale to be $t_v = 0.01$~s. Although variations in GRB spectra down to
milliseconds have been observed, what is important here is the Fourier
strength on the variability, and the choice can be debated. In any
case, when obtaining our final results we will return to this problem
and vary this as well as other critical benchmark parameters over a
wide range.

Long-duration GRB are beamed; their observed isotropically-equivalent
$\gamma$-ray luminosity is on average $L_\gamma =10^{52}$~erg/s. In
the fireball model it results from synchrotron radiation by electrons
accelerated in internal shocks. The spectral photon density of the
burst (GeV${}^{-1}$ cm${}^{-3}$) in the observer's frame can be
adequately parametrized as~\cite{Band:1993eg}
\begin{equation}
n_\gamma \propto 
\begin{cases}(\epsilon/\epsilon_0)^\alpha
e^{-\epsilon/\epsilon_0}&\epsilon<(\alpha-\beta)\epsilon_0\\
(\alpha-\beta)^{\alpha-\beta}e^{\beta-\alpha}(\epsilon/\epsilon_0)^\beta
&\epsilon>(\alpha-\beta)\epsilon_0
\end{cases}
\end{equation}
with $\alpha\simeq-1$, $\beta\simeq-2.2$ and
$\epsilon_0\simeq1$~MeV. The normalization is given by $U_\gamma =
\int {\rm d}\epsilon\,\epsilon n_\gamma(\epsilon) = L_\gamma/4\pi
r_i^2$. The corresponding photon density in the comoving frame is then
given by $n'_\gamma(\epsilon') = n_\gamma(\Gamma_i\epsilon')$.

After Fermi acceleration the electron population follows a power-law
spectrum with minimum energy $E'_{e,{\rm min}} \simeq
(\varepsilon_e/\zeta_e) m_p$ in the comoving frame. Here $\zeta_e<1$
is the fraction of electrons that achieve
equipartition~\cite{Meszaros:2006rc}. This spectrum yields by
synchrotron radiation a photon spectrum peaking at
\begin{equation}\label{eq:synpeak}
  \epsilon_{0} \simeq\Gamma_i \frac{3}{2}\frac{e B'}
  {m^3_e}\left(\frac{\varepsilon_e}{\zeta_e} m_p\right)^2 \simeq
  0.8\,\left( \frac{\varepsilon_{e,-1}^3 \varepsilon_{B,-1}
  L_{\gamma,52}} {\zeta_{e,-1}^4\Gamma_{i,2.5}^4 t_{v,-2}^2}
  \right)^{1/2}\!{\rm MeV}\,,
\end{equation}
where $L_\gamma = L_{\gamma,52}10^{52}$~erg/s, $\varepsilon_{B}
=\varepsilon_{B,-1}0.1$, $\varepsilon_{e} =\varepsilon_{e,-1}0.1$,
$\zeta_{e} =\zeta_{e,-1}0.1$, $\Gamma_{i} = \Gamma_{i,2.5}10^{2.5}$
and $t_{v,-2} = t_{v,-2}0.01$~s. This is close to the observed peak of
the GRB burst spectrum at $\mathcal{O}({\rm MeV})$.

This concludes our description of the GRB fireball and its
electromagnetic spectrum. There is nothing new here. We will next
discuss the production of cosmic rays and neutrinos by applying the
fireball phenomenology just described.

\section{Cosmic Ray and Neutrino Emission}

Although it is straightforward to argue that GRB fireballs represent
an environment that can yield CRs of very high energy, it is important
to take into account that the acceleration competes against energy
loss due to synchrotron radiation and pion production. In other words,
the fireball phenomenology is subject to the conditions that (i) the
time to accelerate protons to the highest energy does not exceed the
lifetime of the fireball and (ii) that the energy gained is not lost
to synchrotron radiation and pion production. We discuss these
constraints sequentially.

For efficient acceleration of UHE CR protons their gyroradius must be
contained within the acceleration region which is related to the size
of the shock. In the comoving frame the Larmor radius of a proton is
$r'_{\rm L} = E'/eB'$ and the corresponding acceleration time
$c/r'_{\rm L}$ is given by $t'_{\rm acc} = \eta r'_L$ with
$\eta\gtrsim1$ or
\begin{equation}
t'_{\rm acc} \simeq 8\times10^{10}\left(\frac{\eta^2\varepsilon_{e,-1}
E_{p,20.5}^2t_{v,-2}^2\Gamma^4_{i,2.5}}{\varepsilon_{B,-1}
L_{\gamma,52}}\right)^{1/2}{\rm
cm}\,,
\end{equation} 
where $E_p =E_{p,20.5}10^{20.5}$~eV, corresponding to the upper end of
the UHE CR spectrum. The size of the accelerator is set by the size of
the shock, $t'_{\rm dyn} \simeq r_i/2\Gamma_i = t_v\Gamma_i$, or
\begin{equation}\label{eq:tdyn}
t'_{\rm dyn} \simeq 1\times10^{11} \Gamma_{i,2.5}t_{v,-2}\,{\rm cm}\,.
\end{equation}
From this we derive the maximal proton energy in the observer's frame
for which $t'_{\rm dyn}=t'_{\rm acc}$,\
\begin{equation}\label{eq:Epmax1}
 E_{p,\rm max} \simeq 4\times 10^{20} \left( \frac{\varepsilon_{B,-1}
 L_{\gamma,52}}{\eta^2\varepsilon_{e,-1} \Gamma_{i,2.5}^{2}}
 \right)^{1/2}\!\!{\rm eV}
\end{equation}
 
We next consider the energy losses that compete with the
acceleration. In the comoving frame the time scale associated with
synchrotron radiation is $t'_{\rm sync} = 9\pi m^4/E'e^4B'^2$ or
\begin{align}\label{eq:tsync}
t'_{\rm sync} 
\simeq 7\times10^{10}\left(\frac{\varepsilon_{e,-1}
\Gamma^7_{i,2.5}t_{v,-2}^2}{\varepsilon_{B,-1}E_{p,20.5}
L_{\gamma,52}}\right) {\rm cm}\,.
\end{align}
The maximal proton energy is reduced when $t'_{\rm syn}=t'_{\rm acc}$ or is smaller; in this case, using (\ref{eq:Epmax1})
\begin{equation}\label{eq:Epmax2}
  E_{p,\rm max} \simeq 3\times 10^{20} \left( \frac{\varepsilon_{e,-1} \Gamma_{i,2.5}^{10} t_{v,-2}^2}{\eta^2\varepsilon_{B,-1} L_{\gamma,52}} \right)^{1/4}\!\!{\rm eV}\,.
\end{equation}

The photo-pion energy loss rate is determined by the $p\gamma$ cross
section, the photon density $n_{\gamma}$ and the average energy loss
of the protons $\langle x_{p\gamma}\rangle$ in each interaction. In
the $\Delta^+$-resonance approximation
\begin{equation}
t'^{-1}_{\Delta^+} \simeq \sigma_\Delta\langle
x_{p\to\Delta^+}\rangle\Gamma_{\Delta^+} \frac{\pi}{2}\frac{m_p}{E'_p}
n_\gamma\left(\frac{\Gamma_im^2_{\Delta^+}}{4E'_p}\right)\,,
\end{equation}
where $\Gamma_{\Delta^+} \simeq 120$~MeV, $m_{\Delta^+} = 1232$~MeV
and $\sigma_{\Delta^+} \simeq 420\,\mu$b and $\langle
x_{p\to\Delta^+}\rangle\simeq 0.2$.  Hence, for a spectral slope
$\alpha\simeq-1$ the optical depth becomes constant above a proton
energy
\begin{equation}\label{eq:Epbreak}
E_{p,{\rm b}} \simeq 0.4\frac{\Gamma_i^2}{\epsilon_{0}}{\rm GeV}^2
\simeq 4\times10^{16}\frac{\Gamma_{i,2.5}^2}{\epsilon_{0,6}}{\rm eV}\,,
\end{equation}
with $\epsilon_0 = \epsilon_{0,6}$~MeV.
Photo-pion production therefore introduces a break in the neutron
production spectrum at (\ref{eq:Epbreak}). Above this energy the
energy loss length in the comoving frame is
\begin{equation}\label{eq:tdelta}
t'_{\Delta} \simeq 4\times10^{12} \left(\frac{\epsilon_{0,6}\Gamma^5_{i,2.5}t_{v,-2}^2}{L_{\gamma, 52}}\right){\rm cm}\,,
\end{equation}
and the maximal proton energy satisfying $t'_{\Delta} = t'_{\rm acc}$ is
\begin{equation}\label{eq:Epmax3}
  E_{p,\rm max} \simeq 2\times 10^{22} \left( \frac{\varepsilon_{B,-1}
  \Gamma_{i,2.5}^{6} t_{v,-2}^2}{\eta^2\varepsilon_{e,-1}L_{\gamma,52}
  } \right)^{1/2}\!\!{\rm eV}\,.
\end{equation}
For GRB fireball parameters considered in this work the scale of
photo-pion losses (\ref{eq:tdelta}) is always larger than the
synchrotron scale (\ref{eq:tsync}) or the dynamical scale
(\ref{eq:tdyn}). The maximal proton energy is hence given by the
smaller of Eqs.~(\ref{eq:Epmax1}) and (\ref{eq:Epmax2}).

There is an additional consideration as mentioned in the introduction:
because protons are magnetically coupled to the expanding fireball
they will lose energy adiabatically if they remain confined in the
expanding shock. Adiabatic cooling, however, is not important for neutrons that are
produced at scales $\langle x_{p\to\Delta}\rangle t'_\Delta$ and can
escape the source~\cite{Rachen:1998ir}.

We will assume that the spectrum of observable CRs results from
neutrons escaping the source. The competition between various energy loss mechanisms in internal shocks can introduce various features in the injection spectrum of individual GRBs~\cite{Protheroe:1998pj}. Within our approximation of a homogeneous source distribution we assume that the {\it effective} CR emission rate can be approximated as
\begin{equation}\label{eq:Qp}
Q_{\rm CR}(E) \simeq Q_0 \frac{(E/E_{p,{\rm
b}})^{-\gamma}}{1+(E/E_{p,{\rm b}})^{\beta-\alpha}}e^{-E/E_{p,{\rm
max}}}\,.
\end{equation}
For $E\gg E_{p,{\rm b}}$, this reduces to the typical $Q_0
E^{-\gamma}\exp(-E/E_{p,{\rm max}})$ approximation of the CR injection
spectrum that we are going to test against HiRes data in the
following. If energy loss of pions and muons prior to decay were
negligible, we can relate the neutrino (per flavor) and CR neutron
emission rates as
\begin{equation}
Q^0_{\nu}(E_\nu) \simeq \frac{1}{\epsilon}Q_{\rm CR}(E_\nu/\epsilon)\,,
\end{equation}
where $\epsilon = \langle E_\nu/E_n\rangle \simeq 0.06$. However, synchrotron losses of secondary pions and muons in the background magnetic field are important. With muon and pion lifetimes of $t_\mu^{\rm dec} = 2.2\,\mu{\rm s}$
and \mbox{$t_\pi^{\rm dec} = 26\,{\rm ns}$}, respectively, this will introduce a synchrotron break in the respective spectrum at
\begin{equation}
E'_{\pi/\mu,{\rm s}} = \frac{3}{4}\sqrt{\frac{m_{\pi/\mu}^5}{\pi\alpha^2 B'^2t_{\pi/\mu}^{\rm dec}}}
\end{equation}
The corresponding break in the neutrino spectra from $\pi^+\to\mu^+\nu_\mu$ and $\mu^+\to e^+\bar\nu_\mu\nu_e$, respectively, is $E'_{\nu,{\rm s}}\simeq E'_{\pi/\mu,{\rm s}}/4$ or
\begin{equation}E_{\nu,{\rm s}} = \left( \frac{\varepsilon_{e,-1} \Gamma_{i,2.5}^8 t_{v,-2}^2}     {\varepsilon_{B,-1} L_{\gamma,52}} \right)^{1/2}\!\!\!\!\times\begin{cases} 2\times 10^{17}     ~{\rm eV} &  (\nu_{\mu})\,,\\ 1\times 10^{16} ~{\rm eV} & ({\bar \nu}_{\mu},     \nu_e)\,. \end{cases}
\end{equation}
The total diffuse neutrino flux can hence be approximated as
\begin{equation}
Q_{{\rm all}\,\nu}(E_\nu) =  \sum_\alpha\frac{Q^0_{\nu}(E_\nu)}{1+(E_\nu/E_{\nu_\alpha,{\rm s}})^2}\,,
\end{equation}
where the sum runs over the neutrino flavors $\nu_\alpha = \nu_\mu$, $\bar\nu_\mu$ or $\nu_e$\,.

\section{Diffuse Cosmic Spectra}

For the calculation of the spectrum of UHE CR protons we assume that
the cosmic source distribution is spatially homogeneous and
isotropic. The comoving number density $Y_i = n_i/(1+z)^3$ of nuclei
of type $i$ is then governed by a set of Boltzmann equations of the
form:
\begin{equation}\label{eq:boltzmann}
\dot Y_i = \partial_E(HEY_i) + \partial_E(b_iY_i)-\Gamma_{i}\,Y_i
+\sum_j\int{\rm d} E_j\,\gamma_{j\to i}Y_j+\mathcal{L}_i\,,
\end{equation}
together with the Friedman-Lema\^{\i}tre equations describing the
cosmic expansion rate $H(z)$ as a function of the redshift $z$:
\mbox{$H^2 (z) = H^2_0\,[\Omega_{\rm m}(1 + z)^3 +
\Omega_{\Lambda}]$}, normalized to its present value of $H_0 \sim70$
km\,s$^{-1}$\,Mpc$^{-1}$, in the usual ``concordance model'' dominated
by a cosmological constant with $\Omega_{\Lambda} \sim 0.7$ and a
(cold) matter component, $\Omega_{\rm m} \sim
0.3$~\cite{Nakamura:2010zzi}.  The time-dependence of the redshift is
${\rm d}z = -{\rm d} t\,(1+z)H$. The first and second term in the
r.h.s.~of Eq.~(\ref{eq:boltzmann}) describe continuous energy losses
(CEL) due to red-shift and $e^{+}e^{-}$ pair
production~\cite{Blumenthal:1970nn} on the cosmic photon backgrounds,
respectively. The third and fourth terms describe more general
interactions involving particle losses ($i \to$ anything) with
interaction rate $\Gamma_i$, and particle generation $j\to i$. We use
the Monte Carlo package {\tt SOPHIA}~\cite{Mucke:1999yb} to calculate
$p\gamma$ interaction rates and spectra. The last term,
$\mathcal{L}_i$, accounts for the CR emission rate per co-moving
volume.

We describe the cosmic evolution of the CR sources by the ansatz
$\mathcal{L}_{\rm CR}(z,E) = \mathcal{H}_{\rm GRB}(z)Q_{\rm CR}(E)$.
Given their supernova association it is natural to assume that the
comoving density of GRBs follows the star formation rate (SFR). We will use
in the following the approximation~\cite{Hopkins:2006bw,Yuksel:2008cu}
\begin{equation}
\label{eq:HSFR}
\mathcal{H}_{\rm SFR}(z) = \begin{cases}(1+z)^{3.4}&z<1\,,\\
N_1\,(1+z)^{-0.3}&1<z<4\,,\\N_1\,N_4\,(1+z)^{-3.5}&z>4\,,
\end{cases}
\end{equation}
with normalization factors, $N_1 = 2^{3.7}$ and $N_4 = 5^{3.2}$. It
has been suggested that the GRB population may not directly follow the
SFR and may have been stronger in the past. This possibility is interesting 
also in the context of UHE CR models: a strong evolution of CR proton sources 
can account for the spectrum of extra-galactic UHE CRs even below the 
ankle~\cite{Berezinsky:2002nc}. These ``low-crossover'' models predict 
significantly larger neutrino fluxes from the sources of CR protons and are 
already limited by upper bounds on diffuse neutrino 
fluxes~\cite{Ahlers:2005sn,Ahlers:2009rf}. For the illustration of the effect 
of a strong GRB rate at high redshift we will use in the following a comparatively 
strong evolution of the from~\cite{Yuksel:2006qb}
\begin{equation}\label{eq:Hstrong}
\mathcal{H}_{\rm strong}(z) = (1+z)^{1.4}\, \mathcal{H}_{\rm SFR}(z)\,.
\end{equation}
This is consistent with a more recent study~\cite{Kistler:2009mv}. However, 
the sample of high-redshift GRBs where the enhanced evolution (\ref{eq:Hstrong}) 
becomes strongest is small and plagued by systematics. It has also been argued 
that the evolution at low redshift is consistent with the SFR whereas at high 
redshift it can be approximated as an almost constant 
rate~\cite{Le:2006pt,Guetta:2007zz,Wanderman:2009es}.

The density of CR sources at high redshift has two effects on the
analysis. Firstly, a stronger evolution of the CR sources requires in
general lower values of the power-law index $\gamma$ to reproduce the
CR data. Secondly, a higher density of sources in the past tend to
produce larger energy densities of secondary neutrinos and
$\gamma$-rays. However, we will see later on that the contribution 
of UHE CR proton sources to the diffuse extra-galactic $\gamma$-ray 
background is already close to maximal if we assume a source evolution 
following the SFR. A stronger evolution as in Eq.~(\ref{eq:Hstrong}) 
can not significantly enhance the prompt neutrino fluxes without violating 
the $\gamma$-ray bound. We will estimate the effect of source evolution on 
the bolometric electro-magnetic energy density in the following.

Particles with electromagnetic (EM) interactions produced in association 
with the cosmic rays, $\gamma$-rays, electrons and positrons, will 
cascade in the universal photon background and magnetic fields on
time-scales much shorter than their production rates. The relevant
processes with background photons are inverse Compton scattering (ICS), $e^\pm+\gamma_{\rm
  bgr}\to e^\pm+\gamma$, pair production (PP), $\gamma+\gamma_{\rm
  bgr}\to e^++e^-$, double pair production (DPP) $\gamma+\gamma_{\rm
  bgr}\to e^++e^-+e^++e^-$, and triple pair production (TPP),
$e^\pm+\gamma_{\rm bgr}\to
e^\pm+e^++e^-$~\cite{Blumenthal:1970nn,Blumenthal:1970gc}. High energy
electrons and positrons can also lose energy via synchrotron radiation
on the intergalactic magnetic field with strength limited
to be $\sim 10^{-9}$G~\cite{Kronberg:1993vk}. 

The evolution of the diffuse $\gamma$-ray and $e^\pm$ spectra follows
the Boltzmann equations~(\ref{eq:boltzmann}). For recent studies
see~\cite{Berezinsky:2010xa,Ahlers:2010fw}. After cascading the EM
flux accumulates into $\gamma$-rays of GeV-TeV energy with a
characteristic and essentially universal spectrum. Its normalization
is determined by the total energy density of EM radiation from the
propagation loss of CR nuclei. For this purpose, we define the
comoving energy density at redshift $z$ as
\begin{equation}
\label{eq:omegacasz}
\omega_{\rm cas}(z) \equiv \int {\rm d} E\,E\!\sum\limits_{i=\gamma,e^\pm}\!Y_i(z,E)\,,
\end{equation}
which follows the evolution equation
\begin{equation}
\dot\omega_{\rm cas} + H\omega_{\rm cas} = \int{\rm d} E\, b(z,E)Y_{\rm CR}(z,E)\,.
\end{equation}
The energy loss rate $b\equiv{\rm d}E/{\rm d}t$ was already
defined by Eq.~(\ref{eq:boltzmann}), but here comprises the combined
energy loss of CR protons into EM radiation by Bethe-Heitler
production and photo-pion interactions.  The energy density (eV
cm${}^{-3}$) of the electromagnetic background observed today is
therefore given by
\begin{equation}
\label{eq:omegacas}
\omega_{\rm cas}  = \int{\rm
d}t\int{\rm d}E \,\frac{b(z,E)}{(1+z)}\,Y_{\rm CR}(z,E)\,.
\end{equation}
Typically the energy density (\ref{eq:omegacasz}) today obtained
by a detailed calculation of the EM spectra agrees with the (quicker)
bolometric calculation (\ref{eq:omegacas}) within a few percent and we
will use the latter in our statistical analysis.

\begin{figure}[t]
\centering
\includegraphics[width = 0.85\linewidth]{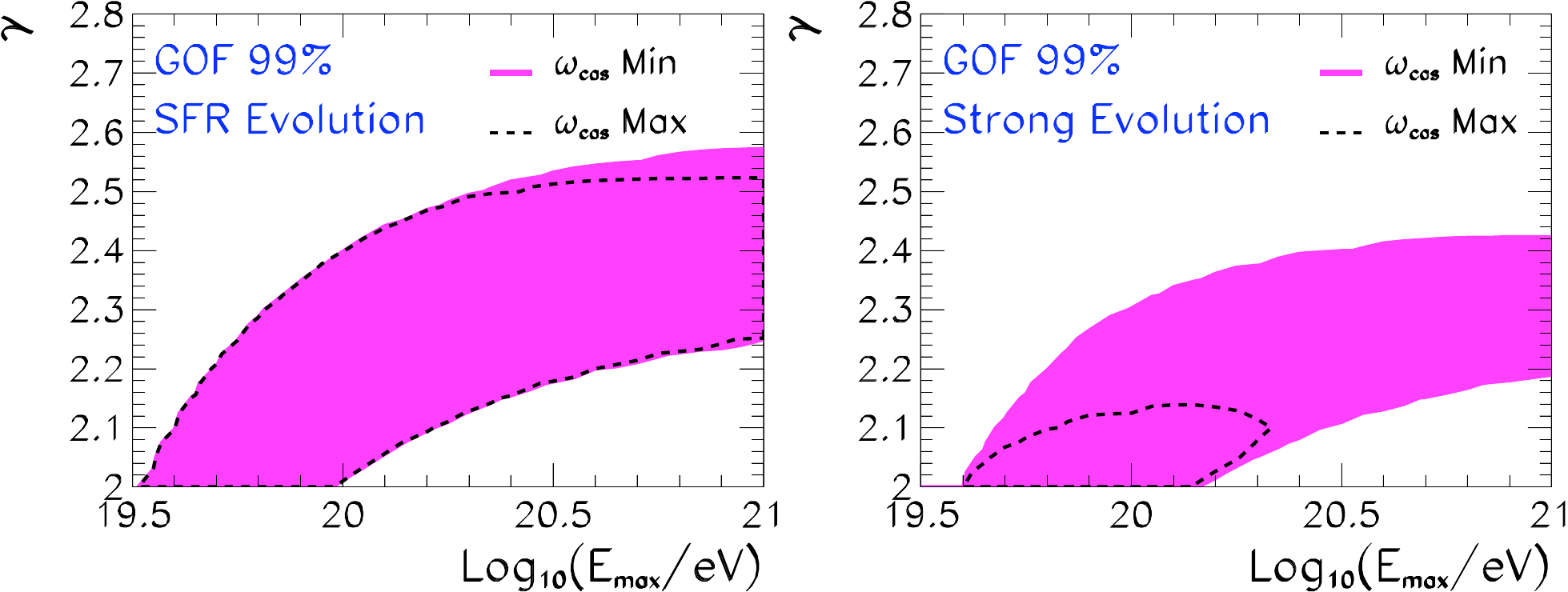}
\caption[]{Results of the goodness of fit test of the HiRes
data~\cite{Abbasi:2007sv}. We show the 99\% (magenta) confidence
levels of the injection index $\gamma$ and the maximal energy $E_{\rm
max}$. In this analysis we consider only minimal $\gamma$-ray
production coming from Bethe-Heitler pair production and photo-pion
production of CR protons during propagation (``Min''). The dashed line
(``Max'') assumes that also $\gamma$-rays from $\pi^0$ production
survive the internal shock environment and cascade during propagation
in the inter-galactic background.}\label{fig:gof}
\end{figure}

The influence of cosmic evolution on the energy density of the cascade
can be estimated in the following way. The UHE CR interactions with
background photons are rapid compared to cosmic time-scales. The
energy threshold of these processes scales with redshift as $E_{\rm
th}/(1+z)$ where $E_{\rm th}$ is the (effective) threshold today. We
can therefore approximate the evolution of the energy density of the
secondaries as
\begin{equation}
\dot\omega_{\rm cas} + H\omega_{\rm cas} 
\simeq \eta_{\rm cas}\mathcal{H}(z)\!\!\!\!\!\!
\int\limits_{E_{\rm th}/(1+z)}\!\!\!\!\!\!{\rm d} E\,E\,Q_{\rm CR}(E)\,,
\end{equation}
where $\eta_{\rm cas}$ denotes the energy fraction of the CR
luminosity converted to the electromagnetic cascade. Assuming a
power-law injection $Q_{\rm CR}(E)\propto E^{-\gamma}$ with
sufficiently large cutoff $E_{\rm max}\gg E_{\rm th}$ we obtain that
strong cosmic evolution (\ref{eq:Hstrong}) enhances the diffuse
$\gamma$-spectrum as $\omega_{\rm cas} \propto \int{\rm
d}t(1+z)^{1.4+\gamma-2}$.  For the proton spectrum $\gamma\simeq2.3$
this corresponds to a relative increase of $\sim4$, which agrees with
numerical results. 

Photo-hadronic interactions in internal shocks produce EM radiation
from the decay of pions. Most of this additional EM radiation is
expected to cascade in the high background densities of $\gamma$-rays
or $e^\pm$ and strong magnetic fields of the fireball and should
eventually contribute to the luminosity of the burst. However, it is
conceivable that part of it decouples from the fireball to contribute
an additional source term $Q_{\rm EM}$~\cite{Razzaque:2004cx} to the
diffuse $\gamma$-ray background. Assuming energy conservation in the
cascade this additional contribution is smaller than
\begin{equation}\label{eq:cassource}
\omega_{\rm cas, source} = \frac{\xi}{H_0}\int{\rm d}E\, E\, Q_{\rm
EM}(E)
\end{equation}
with $\xi/H_0 \equiv \int{\rm d}t\mathcal{H}(z)/(1+z)$. Assuming the
GRB evolution (\ref{eq:HSFR}) and (\ref{eq:Hstrong}) this gives
$\xi_{\rm SFR} \simeq 2.4$ and $\xi_{\rm strong} \simeq 7.3$,
respectively.

In the following we will assume two limiting cases for the
contribution of GRBs to the diffuse GeV-TeV background. The minimal
model (``Min'') assumes that the only contribution to the cascade
results from the propagation from CR protons. In a maximal model
(``Max'') we assume that additionally all $\gamma$-rays from neutral
pion production in the GRB contribute maximally to the cascade in the
form of the term (\ref{eq:cassource}) with $Q_{\rm EM} = Q_{\pi^0}$.

\begin{figure}[t]
\centering \includegraphics[width = 0.85\linewidth]{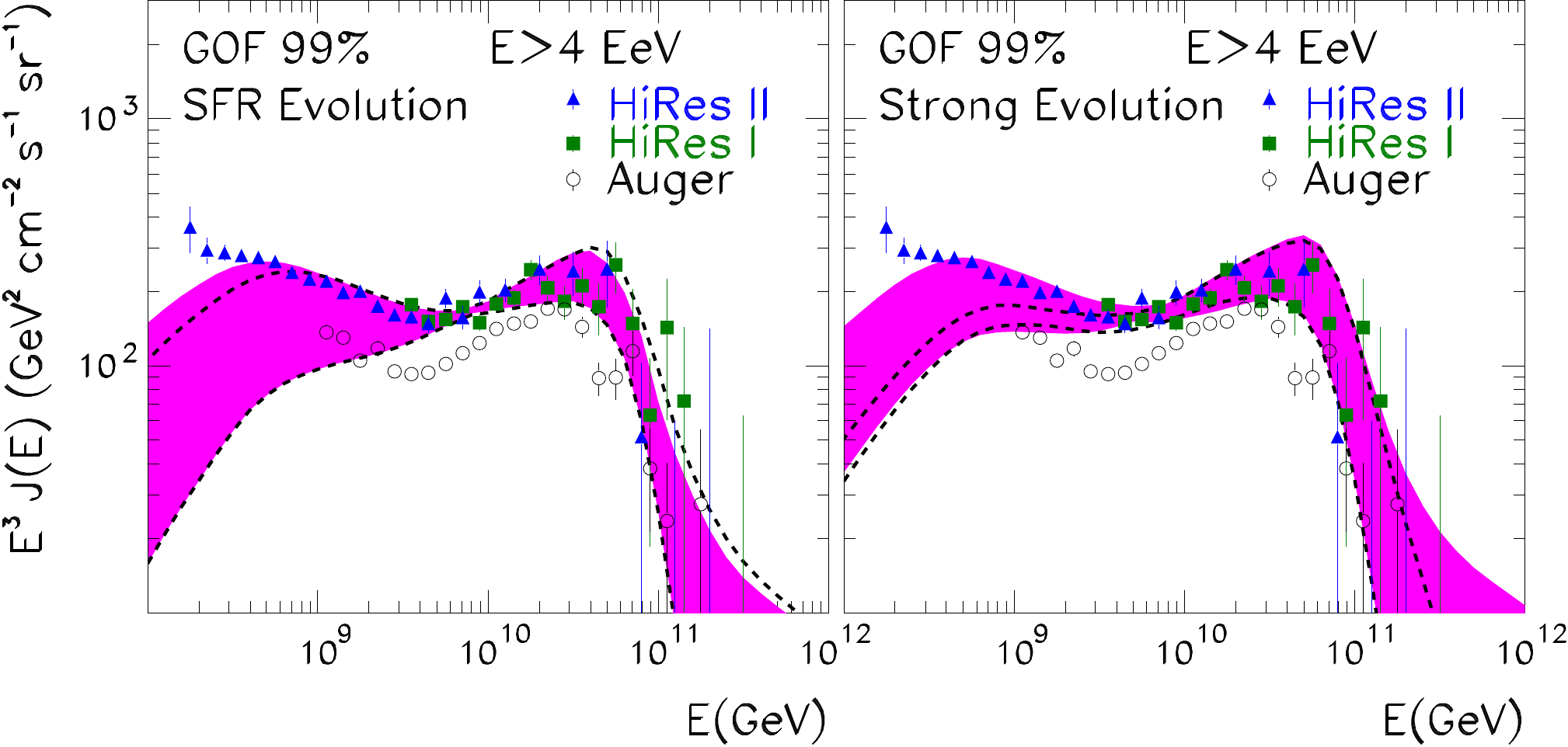}
\caption[]{The allowed proton range of proton flux at the 99\%
confidence level. The color-coding is as in Fig.~\ref{fig:gof}. Each
fit of the proton spectrum is marginalized with respect to the
experimental energy uncertainty and we show the shifted predictions in
comparison to the HiRes central values~\cite{Abbasi:2007sv}.  For
comparison we also show the Auger data~\cite{Abraham:2010mj} which has
{\em not} been included in the fit.}\label{fig:pspectra}
\end{figure}

\section{Goodness of Fit Test}

In this section we present the details and results of our
statistical analysis. We perform a goodness of fit (GOF) test of the
compatibility of cosmic ray data with a given model, characterized by
the spectral index $\gamma$ and the maximal energy $E_{\rm max}$ of
the injected cosmic ray and by the two evolution models
(\ref{eq:HSFR}) and (\ref{eq:Hstrong}). We use CR data from HiRes I
and II \cite{Abbasi:2007sv} above the ``ankle'' at 4~EeV. We
additionally impose the requirement that the electromagnetic flux
accompanying the cosmic rays does not exceed the Fermi-LAT
measurements of the diffuse extra-galactic $\gamma$-ray background in
the GeV-TeV energy range.

Given the acceptance $A_i$ (in units of area per unit time per unit
solid angle) of the experiment in bin $i$ centered at energy $E_i$
with bin width $\Delta_i$ and energy scale uncertainty $\sigma_{E_s}$,
the number of events expected in the bin is
\begin{equation}
N_i(\gamma,E_{\rm max},{\cal N},\delta)= A_i\!\!\!\!
\int\limits_{E_i(1+\delta)-\Delta_i/2}^{E_i(1+\delta)+\Delta_i/2}\!\!\!\!
{\rm d}E\, J^p_{{\cal N},\gamma,E_{\rm max}}(E)\,,
\label{eq:nev}
\end{equation}
where $J^p_{{\cal N},\gamma,E_{\rm max}}(E)=n_p(0,E)/4\pi$ is the
proton flux arriving at the detector. It is determined by the source
luminosity of Eq.~\eqref{eq:Qp} with $E_{p,b}\ll E_{\rm min} = 4$~EeV
and cosmic evolution given by Eqs.~\eqref{eq:HSFR}
or~\eqref{eq:Hstrong}.  The parameter $\delta$ in Eq.~\eqref{eq:nev}
is a fractional energy-scale shift that takes into account the
uncertainty in the energy-scale and ${\cal N}$ is the normalization of
the proton source luminosity.

The probability distribution of events in the $i$-th bin follows a
Poisson ditribution with mean $N_i$.  Correspondingly, the
$r$-dimensional probability distribution for a set of non-negative
integer numbers ${\vec k}=\{k_1,...k_r\}$, $P_{\vec k}(n,\gamma,{\cal
N},\delta)$ is just the product of the individual Poisson
distributions with $r$ the number of bins with $E_i\geq E_{\rm
min}$. Given a model, the experimental result ${\vec N^{\rm
exp}}=\{N^{\rm exp}_1,...,N^{\rm exp}_r\}$ has a probability $P_{\vec
N^{\rm exp}}(\gamma,E_{\rm max},{\cal N},\delta)$ which, after
marginalizing over the uncertainty in the energy scale and
normalization, is given by
\begin{equation}
P_{\rm exp}(\gamma,E_{\rm max})={\rm max}_{\delta,{\cal N}}\left(
P_{\vec N^{\rm exp}}(\gamma,E_{\rm max},{\cal N},\delta)\right)\,.
\label{eq:margi}
\end{equation}
Here the maximization is for some prior for $\delta$ and ${\cal N}$.
As a prior the energy shift $\delta$ we used a top hat
of width $\sigma_{E_s}$.

For ${\cal N}$ we impose two priors. The first is associated with the
upper bound on the total EM of Eq.~\eqref{eq:omegacas} that should not
exceed the Fermi-LAT measurements~\cite{Abdo:2010nz}, or following
Ref.~\cite{Berezinsky:2010xa}
\begin{equation}
w_{\rm cas}({\cal N} ,\gamma,E_{\rm max})
\leq 5.8\times 10^{-7}\; {\rm eV}/{\rm cm}^3\,. 
\label{eq:fermilat}
\end{equation}
The second prior on the normalization is imposed by requiring that the
proton spectra do not exceed the HiRes I and II data below $E_{\rm
min}$ by more than three standard deviations.

The marginalization in Eq.~(\ref{eq:margi}) also determines ${\cal
  N}_{\rm best}$ and $\delta_{\rm best}$ for the model, which are the
values of the energy shift and normalization that yield the best
description of the experimental CR data, subject to the constraint
imposed by the Fermi-LAT measurement.

In the end the model is compatible with the experimental 
results at a given GOF if
\begin{equation}
\sum\limits_{P_{\vec k}\,>\,P_{\rm exp}} 
P_{\vec k}(\gamma,E_{\rm max},{\cal N}_{\rm best},\delta_{\rm best})
\leq {\rm GOF}\,.
\end{equation}
Technically, this calculation is performed by generating a large
  number $N_{\rm rep}$ of replica experiments following the
  probability distribution $P_{\vec k}$ and by imposing that the
  fraction $F$ with $P_{\vec k}>P_{\rm exp}$ satisfy $F\leq{\rm GOF}$.

With this method we determine the range of values of $(\gamma,E_{\rm
max})$ that are compatible with the HiRes I and HiRes II
data~\cite{Abbasi:2007sv}.  We show in Fig.~\ref{fig:gof} the regions
with GOF 99\% for the two evolution models (\ref{eq:HSFR}) (left
panel) and (\ref{eq:Hstrong}) (right panel).  In order to illustrate
the impact of the constraint imposed by the Fermi-LAT measurements, we
also show the GOF regions without imposing it. The region bounded by
the dotted line shows the reduced parameter space resulting from the
condition that secondary EM radiation from proton propagation does not
exceed the Fermi-LAT measurement (``Min''). This bound can become even
stronger if we consider UHE $\gamma$-ray emission from $\pi^0$ decay
in the source with diffuse energy density (\ref{eq:cassource}) as
indicated by the region bounded by the dashed line
(``Max''). Figure~\ref{fig:pspectra} shows the range of the
corresponding proton fits to the data. 

We have not included in the analysis the results from the Auger
Collaboration \cite{Abraham:2008ru,Abraham:2010mj},  which are shown
in Fig.\ref{fig:pspectra} for illustration only.  As
described in Refs.~\cite{Abraham:2008ru,Abraham:2010mj}, besides the
energy scale uncertainty there is also an (energy-dependent) energy
resolution uncertainty which implies that bin-to-bin migrations
influence the reconstruction of the flux and spectral shape. Since the
form of the corresponding error matrix is not public, this data
\cite{Abraham:2008ru,Abraham:2010mj} cannot be analysed outside the
Auger Collaboration.

\begin{figure}[p]
\centering
\includegraphics[width =0.9\linewidth]{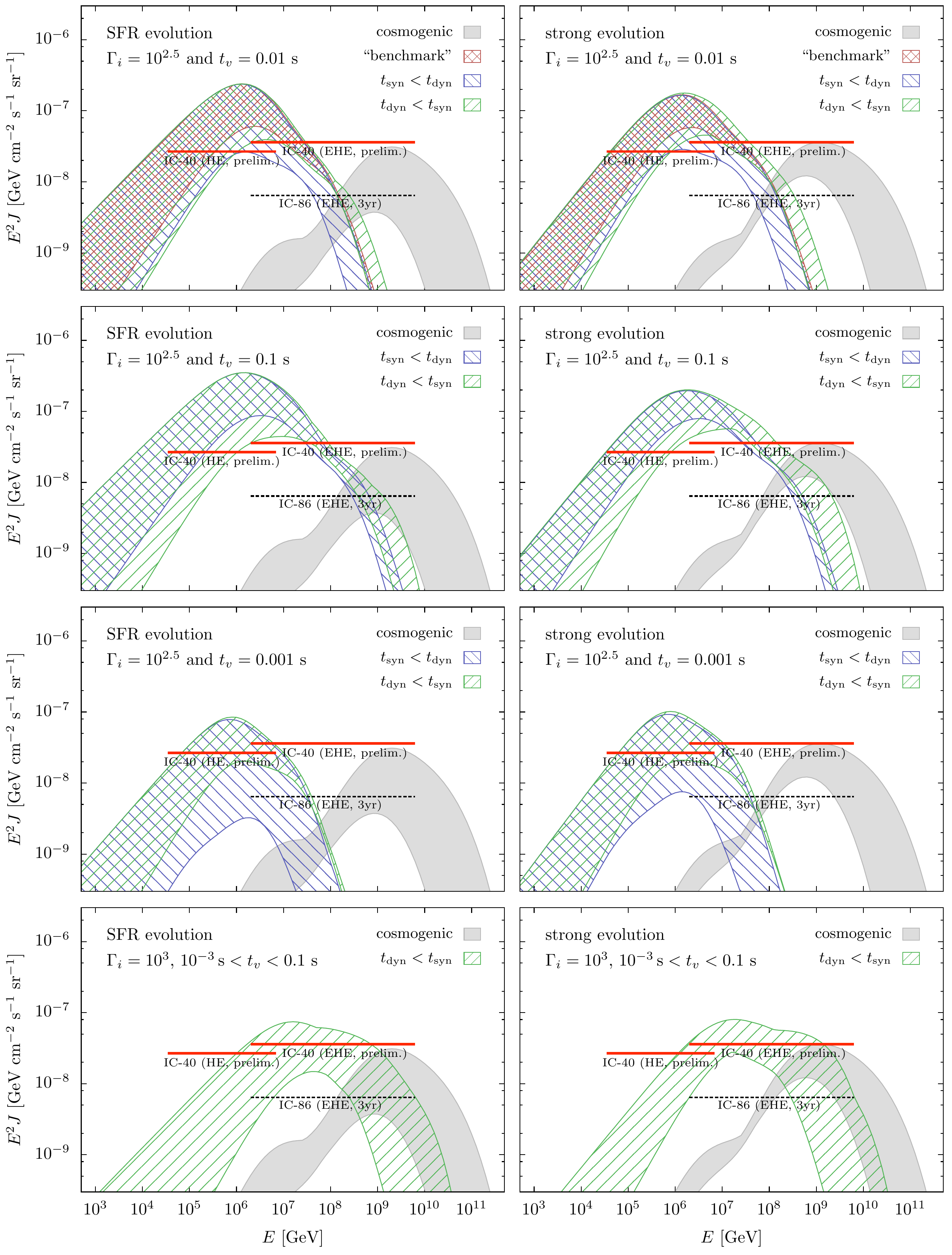}
\caption[]{Prompt neutrino spectra for SFR (left) and strong (right)
evolution corresponding to the 99\% C.L.~of the ``Min'' model shown 
in Fig.~\ref{fig:gof} and assuming the luminosity range
$0.1<(\varepsilon_B/\varepsilon_e)L_{\gamma,52}<10$. We show the
prompt spectra separately for the branches $t_{\rm dyn}<t_{\rm syn}$
(green right-hatched) and $t_{\rm syn}<t_{\rm dyn}$ (blue
left-hatched). The preliminary IceCube limits~\cite{IceCubeEHE,Abbasi:2011ji} (90\% C.L.) on the total
neutrino flux from the analysis of high-energy (HE) and extremely-high energy (EHE) 
neutrinos with the 40 string sub-array (IC-40) assume 1:1:1 flavor
composition after oscillation. We also show the sensitivity of the
full IceCube detector (IC-86) to EHE neutrinos after three years
of observation. The gray solid area shows the range of GZK neutrinos
expected at the 99\% C.L.}\label{fig:nuspectra}
\end{figure}

\section{Results and Their Dependence on Model Parameters}

Our final results are summarized in Fig.~\ref{fig:nuspectra} that
shows the range of prompt neutrino spectra corresponding to the UHE CR
proton spectra at the 99\% C.L.~of the GOF test with the HiRes data
assuming SFR evolution (left panels) and the stronger evolution of
Eq.~(\ref{eq:Hstrong}) (right panels). We have assumed that proton
acceleration in the shocks is efficient ($\eta\simeq1$) and have fixed
the maximum $\epsilon_0$ of the prompt $\gamma$-ray emission
(Eq.~(\ref{eq:synpeak})) at 1~MeV. For each GRB model the maximum
proton energy $E_{\rm max}$ corresponds to the smaller of
Eq.~(\ref{eq:Epmax1}) or (\ref{eq:Epmax2}), {\it i.e.}~$t_{\rm dyn} <
t_{\rm syn}$ or $t_{\rm syn}< t_{\rm dyn}$, respectively. For given
values of $E_{\rm max}$, $\Gamma_i$ and $t_v$ we can in general derive
two solutions of $(\varepsilon_B/\varepsilon_e)L_{\gamma}$, which we
allow to vary within the range
$0.1<(\varepsilon_B/\varepsilon_e)L_{\gamma,52}<10$. The corresponding
branches are indicated in the plots.

The upper three rows of Fig.~\ref{fig:nuspectra} shows prompt neutrino
spectra with a Doppler factor $\Gamma_i = 10^{2.5}$ of the internal
shock. The top panels show the results for the variation
time-scale $t_v=0.01$. This includes the range of neutrino fluxes for
typical ``benchmark'' values of the GRB environment with $L_{\gamma} =
10^{52}$~erg/s and $E_{\rm max}=10^{20.5}$~eV, which is shown separately
as the cross-hatched area. The range of prompt neutrino spectra for
the benchmark model corresponds to the range of the power-law index
$\gamma$ at the 99\% C.L.~and the normalization of each particular
model. Note, that for both cases, SFR or strong evolution, the range
of the prompt neutrino flux exceeds the present limits of IC-40.

The energy density of the prompt neutrino spectrum peaks at an energy
$E\simeq \epsilon E_{p,b}\simeq1$~PeV and is marginally consistent
with recent upper limits on the diffuse neutrino flux by
IC-40~\cite{IceCubeEHE,Abbasi:2011ji}. Only for very short variation time-scales of
$t_v\simeq10^{-3}$~s (third row of Fig.~\ref{fig:nuspectra}) and
for synchrotron-loss-dominated GRBs ($t_{\rm syn} < t_{\rm dyn}$) can
the predicted flux avoid present upper limits. Note that since
$E_{p,b}\propto\Gamma_i^2/\epsilon_0$ we can shift the peak to higher
energies assuming a stronger Doppler boost of the internal
shocks. However, the condition $E_{p,b}\lesssim E_{\rm min}$ imposed
for the fit requires that $\Gamma_i\lesssim10^{3.5}$ and hence
$\epsilon E_{p,b}\lesssim100$~PeV.

The bottom row of Fig.~\ref{fig:nuspectra} we show the prompt neutrino
spectra for $\Gamma_i=10^3$ and the full range of variability
$10^{-3}<t_v/{\rm s}<0.1$. Due to the strong dependence of the
synchrotron time scale on the Doppler factor, $t'_{\rm syn}\propto
\Gamma_i^7$, the GRB models at the 99\% C.L.~correspond to proton
sources where the maximal energy is limited by the dynamical time. The
range of the flux is consistent with present IC-40 limits. Note,
however, that the opacity of of the GRBs for UHE $p\gamma$
interactions and subsequent emission of CR neutrons is very sensitive
to the Doppler factor,
\begin{equation}
\tau_{p\gamma} \simeq \frac{t'_{\rm dyn}}{t'_\Delta\langle
x_{p\to\Delta}\rangle}\simeq0.1\left(\frac{L_{\gamma,52}}
{\Gamma_{i,2.5}^4t_{v,-2}\epsilon_{0,6}}\right)\,.
\end{equation}
Increasing the average Doppler factor of the GRB from $10^{2.5}$ to
$10^3$ reduces the efficiency of UHE CR emission by two orders of
magnitude. It is hence more likely that the successful candidates of
CR sources are GRBs with lower $\Gamma_i$.

The range of prompt neutrino spectra shown in Fig.~\ref{fig:nuspectra} 
spans about two orders of magnitude, a range larger than a previous
study in Ref.~\cite{Guetta:2001cd}. The difference of the results can
be traced back to different assumptions. Firstly, we here assumed that
only neutrons can escape the GRB shock environment and form the
spectrum of UHE CRs. Hence, both, CR and neutrino spectra depend on
the opacity of the source. Secondly, we apply a statistical fit to the
data assuming a wide range of spectral indices. Statistically allowed
CR spectra with a steep injection index $\gamma\gg2$ correspond to GRB
models with a higher bolometric energy density and hence larger prompt
neutrino emission. For large maximal proton energies, $E_{\rm max}\gg 
10^{19}$~eV, we can approximate the total local power density in UHE CRs 
derived from the best fit to the HiRes data as
\begin{equation}
\dot \varepsilon_{\rm CR} \equiv \int\limits_{E_{p,{\rm b}}}{\rm d}EE
{\mathcal L}_{\rm CR}(0,E) \simeq (1-2)\times 10^{44}~{\rm erg}\,{\rm Mpc}^{-3}
\,{\rm yr}^{-1}\times\begin{cases}\frac{1}{\gamma-2}\left(
\frac{10^{19}{\rm eV}}{E_{p,{\rm b}}}\right)^{\gamma-2}&\gamma>2\,,\\
\ln\left(\frac{E_{\rm max}}{E_{p,{\rm b}}}\right)&\gamma=2\,.\end{cases}
\end{equation}
This value is consistent with the original study by Waxman and 
Bahcall~\cite{Waxman:1997ti} assuming a total power density in UHE CRs 
of the order of $4\times10^{44}~{\rm erg}\,{\rm Mpc}^{-3}\,{\rm yr}^{-1}$ 
between $10^{19}$~eV and $10^{21}$~eV assuming a flat spectrum with 
power-law index $\gamma=2$. In our approach we allow steeper injection 
spectra $\gamma\gg2$ extending down to the break energy $E_{p,{\rm b}}\ll 
10^{19}$~eV ({\rm cf}.~Eq.~(\ref{eq:Epbreak})) which is set by the GRB 
fireball environment. For very steep injection spectra with $\gamma\simeq2.6$ 
consistent with our fit at the 99\% C.L.~({\rm cf.}~left panel in 
Fig.~\ref{fig:gof}) and  $E_{p,{\rm b}} = 4\times10^{16}$~eV this corresponds 
to a maximal local power density of the order of $\dot \varepsilon_{\rm CR, {\rm max}} 
\simeq 10^{46}~{\rm erg}\,{\rm Mpc}^{-3}\,{\rm yr}^{-1}$.

These large local power densities of steep UHE CR proton spectra are 
also consistent with the results of Ref.~\cite{Wick:2003ex}. In contrast 
to Ref.~\cite{Wick:2003ex} we include here the limits implied by the 
extra-galactic diffuse $\gamma$-ray background inferred by Fermi LAT. 
This bound has the strongest effect on strong evolution models with 
large power-law index as can be seen in the right panel of 
Fig.~\ref{fig:gof}. The corresponding proton fluxes at the 99\% C.L.~are 
shown in the right panels of Fig.~\ref{fig:pspectra} and the prompt 
neutrino fluxes in the right panels of Fig.~\ref{fig:nuspectra}. 
The range of spectra is comparable to the moderate evolution following 
SFR (left panels) since the contribution of UHE CRs to the diffuse 
extra-galactic $\gamma$-ray background is already close to maximal.

\section{Conclusion}

We have discussed prompt neutrino emission associated with the
production of UHE CR protons in GRBs. Our analysis assumes that UHE
CRs above the ``ankle'' at 4~EeV consists of neutrons emitted from
$p\gamma$ interactions in internal shocks of the GRB fireball
model. We determined the CR emission density by a fit to HiRes data
and used this information to determine the corresponding neutrino
fluxes for a wide range of parameters for the shock environment. We
have also carefully studied the effect of secondary production during
CR propagation in the form of GZK neutrinos as well as the GeV-TeV
$\gamma$-ray background.

The main results are shown in Fig.~\ref{fig:nuspectra}. We have
shown that typical (``benchmark'') fireball environments predict
prompt neutrino fluxes that exceed present diffuse neutrino limits
from IceCube (IC-40) if the associated proton spectrum is fitted to
actual CR data. This remains partially true if we relax the condition
of internal shock parameters. Consistency with the diffuse neutrino
limits requires that the CR acceleration takes place in GRB fireballs
with small internal shock radii (corresponding to variation
time-scales of milli-seconds) and/or large Lorentz boost
$\Gamma_i\simeq1000$. Alternatively, it is possible that UHE CRs at 
the ankle might still receive a significant contribution from galactic 
CRs, which lowers the required power density in extra-galactic CRs 
and the corresponding prompt neutrino spectra from GRB 
sources~\cite{Katz:2008xx}. However, one has to worry that galactic 
CRs at these energies will reveal themselves via an anisotropy toward 
the galactic center. In any case, the sensitivity of the full IceCube observatory
(IC-86) after three years of observation will improve present diffuse
neutrino limits by a factor five and will serve as a crucial test of
the GRB scenario of UHE CRs. 

Finally, we would also like to mention additional constraints of
this CR scenario. A fraction of UHE $\gamma$-rays produced in neutral
pion production may escape the fireball shocks and contribute to the
GeV-TeV background as well. As an estimate, we have tested a
pessimistic scenario (``Max'' model) where all $\gamma$-rays from $\pi^0$ production
escape the source. We found that this additional contribution may
become important in the case of source evolution much stronger than
the SFR ({\it cf.}~dashed regions in Figs.~\ref{fig:gof} and \ref{fig:pspectra}). 
It is also possible to constrain UHE CR emission from
internal shocks of GRBs by the integrated fluence of all bursts as has
been done recently in~\cite{Eichler:2010ky}. On the other hand, the
contribution of the burst spectrum to the diffuse $\gamma$-ray
background from unidentified cosmic GRBs has been discussed
in~\cite{Dermer:2006vd} and has been shown to contribute a relatively
small fraction.
 
\section*{Acknowledgments}
The authors would like to thank Peter Meszaros, Kohta Murase, Eli Waxman 
and Walter Winter for valuable comments on the manuscript.
This work is supported by US National Science Foundation Grant No
PHY-0969739  and by the Research Foundation of SUNY at
Stony Brook. F.H. is supported by
U.S. National Science Foundation-Office of Polar Program,
U.S. National Science Foundation-Physics Division, and the University
of Wisconsin Alumni Research Foundation. M.C.G-G acknowledges further
support from Spanish MICCIN grants 2007-66665-C02-01,
consolider-ingenio 2010 grant CSD2008-0037 and by CUR Generalitat de
Catalunya grant 2009SGR502.

\end{document}